%%%%%%%%%%%%%%%%%%%%%%%%%%%%%%%%%%%%%%%%%%%%%%%%%%%%%%%%%%%%%%%%%%%%%
\input harvmac
\font\cmss=cmss10 \font\cmsss=cmss10 at 7pt
\def\IZ{\relax\ifmmode\mathchoice
{\hbox{\cmss Z\kern-.4em Z}}{\hbox{\cmss Z\kern-.4em Z}}
{\lower.9pt\hbox{\cmsss Z\kern-.4em Z}}
{\lower1.2pt\hbox{\cmsss Z\kern-.4em Z}}\else{\cmss Z\kern-.4em
Z}\fi}
\def\dphi{\tilde\phi_{d-h}}
\def\omegad{\tilde\omega_{d-h-1}}

\def\WW{W_\alpha W^\alpha}
\def\Dterm#1{\left( #1 \right)_{ D}}
\def\Fterm#1{\left( #1 \right)_{ F}}
\def\d{\delta}

\Title{\vbox{\baselineskip12pt\hbox{CERN-TH/96-109} \hbox{UGVA-DPT
1996/04-923}
\hbox{hepth/9604196 }}}
{\vbox{\centerline{Phases of Antisymmetric Tensor Field Theories }
%\bigskip%\centerline{Duality, Massive Phases and $p$-Branes}
}}
\medskip
\centerline{Fernando Quevedo$^a$ and Carlo A. Trugenberger$^b$
}%\foot{$\dagger$}{Supported by a profil 2 fellowship of the Swiss
% National
%Science Foundation.}
\bigskip\centerline{$^a$ {\it CERN Theory Division, CH-1211, Geneva
23, Switzerland}}
\centerline{$^b$ {\it Department of Theoretical Physics, University
of Geneva}}
\centerline{\it 24, quai E. Ansermet, CH-1211 Geneva 4, Switzerland}
\bigskip\bigskip\bigskip\medskip
\noindent
We study the different phases of field theories of compact
antisymmetric tensors of rank $h-1$
in arbitrary space-time dimensions $D=d+1$.
Starting in a `Coulomb' phase, topological defects of dimension
$d-h-1$ ($(d-h-1)$-branes)
may condense leading to a generalized `confinement'
phase. If the dual theory is also compact the model
may also have a
third, generalized `Higgs' phase, driven by the condensation
of the dual $(h-2)$-branes.
Developing on the work of Julia and Toulouse for ordered solid-state
media,
we obtain the low energy
effective action for
these  phases.
Each phase has two  dual
descriptions in terms of  antisymmetric tensors of different ranks,
which are massless for the Coulomb phase but massive for the
Higgs and confinement phases. We illustrate our prescription in
detail for compact QED in 4D.
Compact QED and  $O(2)$ models in 3D,
as well as  a  periodic scalar field in 2D (strings on a circle),
are also discussed.
In this last case we show how $T$-duality is maintained if one
considers both
worldsheet instantons and their duals.
We also unify various approaches to the problem of the axion mass in
4D string models.
Finally we discuss  possible implications of our results for
non-perturbative issues in string theory.%condensation of $p$-branes
% in any %dimension
%and the possible selection of string
%vacua.
\Date{\vbox{\baselineskip12pt\hbox{CERN-TH/96-109}
\hbox{April 1996}}}

\lref\jt{B. Julia and G. Toulouse, J. Physique, Lett. 40 (1979) 396.}

\lref\klls{R. Kallosh, A. Linde, D. Linde and L. Susskind, Phys. Rev.
D
52 (1995) 912.}

\lref\bdqq{C.P. Burgess, J.-P. Derendinger, F. Quevedo and M.
Quir\'os,
Phys. Lett. B348 (1995) 428.}

\lref\bin{P. Binetruy, M.K. Gaillard and T. Taylor, Nucl. Phys. B455
(1995) 97;
P. Binetruy, F. Pillon, G. Girardi and R. Grimm, hep-th/9603181.}

\lref\poly{A.M. Polyakov, `` Gauge Fields and Strings'', Harwood
Academic Publishers (1987).}

\lref\sath{B. Sathiapalan, Phys. Rev. D35 (1987) 3277;
I. Kogan, JETP Lett. 45 (1987) 709.}

\lref\gk{D.J. Gross and I.Klebanov, Nucl. Phys. B344 (1990) 475.}

\lref\kt{M. Kosterlitz and D. Thouless, J. Phys. C6 (1973) 1181;
V.L. Berezinski, JETP 34 (1972) 610;
J. Villain, J. Phys. C36 (19975) 581.}

\lref\gsw{M. Green, J. Schwarz and E. Witten,
``Superstring Theory'' 2 Vols, Cambridge
University Press, Cambridge, (1987).}

\lref\att{A. Aurilia, Y. Takahashi and P.K. Townsend,
Phys. Lett. 95B (1980) 265; A. Aurilia,
F. Legovini and E. Spalucci, Phys. Lett. B264 (1991) 69.}

\lref\gpr{For a review see: A. Giveon, M. Porrati and E. Rabinovici,
Phys. Rep. 244 (1994) 77;
E. Alvarez, L. Alvarez-Gaum\'e and Y. Lozano, hep-th/9410237.}

\lref\dkl{For a review see: M.J. Duff, R. Khuri and J. Lu, Phys. Rep.
259 (1995) 213.}

\lref\sjo{S.-J. Rey, Phys. Rev. D40 (1989) 3396.}

\lref\sjt{S.-J. Rey, Phys. Rev. D43 (1991) 526.}

\lref\joe{J. Polchinski, hep-th/9510017.}

\lref\topmass{T.L. Curtright and P.G.O. Freund,
{\it Nucl. Phys.} B172 (1980) 413;
T. Curtright, { Phys. Lett.} 165B (1985) 304;
 P.K. Townsend, K. Pilch and P. van Nieuwenhuizen,
{ Phys. Lett.} 136B (1984) 38; S. Deser and
R. Jackiw, { Phys. Lett.} 139B (1984) 371.}

\lref\fintemp{K.H. O'Brien and C.I. Tan, Phys. Rev. D36 (1987) 1184;
E. Alvarez and M.A.R. Osorio, Nucl. Phys. B304 (1988) 327;
J. Atick and E. Witten, Nucl. Phys. B310, (1988) 291.}

\lref\olive{For a recent review see: D. Olive, hep-th/9508089.}

\lref\sw{N. Seiberg and E. Witten, Nucl. Phys. B426  (1994) 19.}

\lref\raj{For a review see: R. Rajaraman, ``Solitons and Instantons",
North-Holland, Amsterdam (1987);
S. Coleman, ``Aspects of Symmetry'', Cambridge University Press,
Cambridge (1985).}

\lref\vy{G. Veneziano and S. Yankielowicz, Phys. Lett. 113B (1982)
231;
C.P. Burgess, J.-P. Derendinger, F. Quevedo and M. Quir\'os,
hep-th/9505171.}

\lref\itz{See for example: C. Itzykson and J.-M. Drouffe,
``Statistical
Field Theory", Cambridge University Press, Cambridge (1989);
J.B. Kogut, Rev. Mod. Phys. 51 (1979) 659.}

\lref\stu{E. C. G. St\"uckelberg, Helv. Phys. Acta 30 (1957) 209.}

\lref\man{G. 't Hooft, in ``High Energy Physics: Proceedings of the
EPS
International Conference, Palermo, June 1975", A. Zichichi ed.,
Editrice
Compositori, Bologna (1976); S. Mandelstam, in ``Extended Systems in
Field
Theory", J. L. Gervais and A. Neveu eds., Phys. Rep. C23 (1976).}

\lref\nash{For a review see: C. Nash and S. Sen, ``Topology and
Geometry for
Physicists", Academic Press, London (1983).}

\lref\pis{R. D. Pisarski, Phys. Rev. D34 (1986) 3851; I. Affleck, J.
Harvey,
L. Palla and G. Semenoff, Nucl. Phys. B328 (1989) 575; M. C.
Diamantini,
P. Sodano and C. A. Trugenberger, Phys. Rev. Lett. 71 (1993) 1969;
F. Schaposnik and J. Edelstein, Nucl. Phys. B425 (1994) 137.}

\lref\godol{P. Goddard and D. Olive, Rep. Progr. in Phys. 41 (1978)
1358.}

\lref\bmk{A. M. Polyakov, Nucl. Phys. B120 (1977) 429; T. Banks, R.
Myerson
and J. Kogut, Nucl. Phys. B129 (1977) 493.}

\lref\ewit{E. Witten, Phys. Lett. Phys. Lett. 86B (1979) 283.}

\lref\jac{For a review see: S. B. Treiman, R. Jackiw, B. Zumino and
E. Witten,
``Current Algebra and Anomalies", World Scientific, Singapore
(1985).}

\lref\caru{C. G. Callan, Phys. Rev. D25 (1982) 2141, Phys. Rev. D26
(1982) 2058, Nucl. Phys. B212 (1983) 391; V. Rubakov, JETP Letters 33
(1981) 644,
Nucl. Phys. B203 (1982) 311; K. Isler, C. Schmid and C. A.
Trugenberger,
Nucl. Phys. B294 (1987) 925.}

\lref\eli{J.V. Jose, L.P. Kadanoff, S. Kirkpatrick and D. R. Nelson,
Phys. Rev. B16 (1977) 1217; S. Elitzur, R. Pearson and J. Shigemitsu,
Phys. Rev D19 (1979) 3698;
D. Horn, M. Weinstein and S. Yankelowicz, Phys. Rev D19 (1979) 3715;
A. Guth, A. Ukawa and P. Windey, Phys. Rev. D21 (1980) 1013.}

\lref\cardy{J. L. Cardy and E. Rabinovici, Nucl. Phys. B205 [FS5]
(1982), 1;
J. L. Cardy, Nucl. Phys. B205 [FS5] (1982) 17; A. Shapere and F.
Wilczek,
Nucl. Phys. B320 (1989) 669.}

\lref\lut{C.P. Burgess and F. Quevedo, Nucl. Phys. B421 (1994) 373;
J. Fr\"ohlich, R. G\"otschmann and P. A. Marchetti, J. Phys A: Math.
Gen. 28 (1995) 1169; C.P. Burgess, C.A. L\"utken and F. Quevedo,
Phys. Lett. B336 (1994) 18.}

\lref\fro{J. Fr\"ohlich and U. Studer, Rev. Mod. Phys. 65 (1993)
733.}

\lref\mavro{A. Kovner and B. Rosenstein, Phys. Rev. B42 (1990) 4748;
G. W. Semenoff and N. Weiss, Phys. Lett. B250 (1990) 117; N. Dorey
and
N. E. Mavromatos, Phys. Lett. B250 (1990) 107, Nucl. Phys. B386
(1992)
614; M. C. Diamantini, P.Sodano and C. A. Trugenberger, preprint
CERN-TH/95-294.}

\lref\dst{M. C. Diamantini, P.Sodano and C. A. Trugenberger, Nucl.
Phys.
B448 (1995) 505.}

\lref\kalb{V.I. Ogievetsky and I.V. Polubarinov, Sov. J. Nucl. Phys.
4 (1967) 156; M. Kalb and P. Ramond, Phys. Rev. D9 (1974) 2273;
Y. Nambu, Phys. Rep. 23 (1976) 250;
D.Z. Freedman and P.K. Townsend, Nucl. Phys. B177 (1981) 282;
J. Thiery-Mieg and L. Beaulieu,
Nucl. Phys. B228 (1983) 259.}

\lref\peccei{For a review see: R.D. Peccei, in ``CP Violation'',
ed. C. Jarlskog, World Scientific, Singapore (1989).}

\lref\polch{For a recent review see: J. Polchinski, hep-th/9511157.}

\lref\claudio{C. Teitelboim, Phys. Lett. 167B (1986) 63.}

\lref\claudiot{C. Teitelboim, Phys. Lett. 167B (1986) 69;
R.I. Nepomechie, Phys. Rev. D31 (1984) 1921.}

\lref\us{F. Quevedo and C.A. Trugenberger, in preparation.}

\lref\polyt{A. Polyakov, in ``Fields, Strings and Critical
Phenomena'',
E. Br\'ezin and J. Zinn-Justin eds., North-Holland, Amsterdam
(1990).}

\lref\paul{P.K. Townsend, hep-th/9507048; K. Becker, M. Becker
and A. Strominger,  hep-th/9507158.}

\lref\orland{R. Savit, Rev. Mod. Phys. 52 (1980) 453;
P. Orland, Nucl. Phys. B205[FS5] (1982) 107.}

\lref\klee{K. Lee, Phys. Rev D48 (1993) 2493.}

\lref\savit{R. Savit, Rev. Mod. Phys. 52 (1980) 453.}

\lref\sdual{A. Font, L.E. Ib\'a\~nez, D. L\"ust, F.
Quevedo, {Phys. Lett.} 249B (1990) 35;
A. Sen, {Phys. Lett.} B303 (1993) 22; Phys. Lett. B329 (1994) 217;
J. Schwarz and A. Sen, { Nucl. Phys.} B411 (1994) 35.
C. Hull and P.K. Townsend, Nucl. Phys. B438 (1995) 109;
E. Witten, Nucl. Phys. B443 (1995) 85.}
%\draft

\noblackbox

\newsec{Introduction}
Antisymmetric tensor theories have been thoroughly studied
during the past years \refs{\kalb,\claudio,\claudiot,\topmass,\att,
\orland}.
They are the natural extension of
free scalar field theories and Abelian gauge theories, and
share some properties which makes them easy to study.
In particular the powerful property of strong--weak coupling
duality \olive, known for electromagnetism and free scalar field
theories, can be easily generalized to antisymmetric tensor
theories in any dimension.

Antisymmetric tensors also appear very naturally in supersymmetric
field  theories and  in  string theories \gsw. They play an important
role in the realization of the various strong-weak
coupling dualities among string theories \refs{\sdual,\polch}.
An antisymmetric tensor of rank $h-1$ couples naturally
to  an elementary extended object of dimension $h-2$, a $(h-2)$-brane
\dkl.
These objects, however, may also appear as solitonic
  excitations \raj\ of an
underlying theory if the antisymmetric tensors are compact variables.

It is well known that the condensation of topological defects may
drive phase transitions, the prototype of this phenomenon being the
vortex-driven Kosterlitz--Thouless transition in two space dimensions
\refs{\kt,\itz}.

When analyzing phase transitions induced by topological defects, two
questions have to be answered. The first is to establish if  a
certain kind of topological defect does indeed condense, and for
which
values of the temperature or the coupling constants this happens. The
second is to establish the nature of the new phase with a finite
condensate of topological defects. In this paper we shall concentrate
on this second aspect for generic antisymmetric tensor field theories
in
$D=d+1$ space-time dimensions.

Nearly twenty years ago, Julia and Toulouse \jt\
tackled this problem in the
framework of ordered solid-state media. They considered models
supporting
stable topological defects, with homotopy group $Z$ \nash,
and characterized
by a length scale $1/M$, where the mass $M$ is considered as a
cut-off
for the low-energy effective field theory.

The idea of Julia and Toulouse is that the condensation of these
topological defects generates new hydrodynamical modes for the
low-energy effective theory: these new modes are essentially the
long wavelength fluctuations of the continuous distribution of
topological defects. Moreover, Julia and Toulouse proposed also a
generic prescription to identify these new modes. However, in the
framework of ordered solid-state media, it is difficult to write
down an action for the phase with a condensate of topological defects
due to the non-linearity of the topological currents, the lack of
relativistic invariance, and the need to introduce dissipation terms.
Here, we will apply and develop the idea of Julia and Toulouse for
generic compact antisymmetric field theories, for which none of the
above problems is present.

In this framework we will show that the Julia--Toulouse prescription
can be made much more precise and that it leads also to an explicit
form for the action in the finite condensate phase. We will also show
that this phase is the natural generalization of the confinement
phase
for a vector gauge field and that the Julia--Toulouse mechanism is
the
exact {\it dual} of the Higgs mechanism in its pristine St\"uckelberg
formulation \stu. Thus the generalized confinement phase for the
original
tensor is equivalent to a generalized Higgs-St\"uckelberg phase for
its dual tensor.
Our results generalize the well-known 't Hooft--Mandelstam
duality \man\ to any $p$-form theory in $D$ space-time dimensions.

We also present several concrete examples, among which a detailed
discussion of 4D compact quantum electrodynamics (QED)
\refs{\bmk\poly}, of
$T$-duality in compactified strings \gpr\ and of the axion mass in 4D
string models \refs{\bdqq,\klls, \bin}.
Finally we discuss possible implications of our results for
non-perturbative issues in string theories.

This paper is organized as follows. In section 2 we present the
original
Julia--Toulouse prescription, while in section 3 we adapt it to
antisymmetric
tensor field theories. In section 4 we discuss the example of compact
QED and we demonstrate the confinement mechanism. Section 5 is
devoted
to the derivation of the duality between the Julia--Toulouse
mechanism and the Higgs--St\"uckelberg mechanism. In section 6 we
discuss various examples. Finally we draw our conclusions in section
7.

\newsec{The Julia--Toulouse prescription}
In this section we shall present the Julia--Toulouse prescription as
originally
developed for ordered solid-state media \jt.

The low-energy excitations of such systems are generically described
by a
field theory for an order parameter. In addition to these propagating
modes,
there are also massive classical solutions describing the {\it
topological
defects} of the medium.

We thus consider a generic $(d+1)$-dimensional
field theory with symmetry
group $G$ spontaneously broken down to $H$. The homotopy groups
$\Pi _h \left( G/H \right) $ classify the topological defects that
can
arise in this theory \nash. A non-trivial $\Pi _h \left( G/H \right)
$ for
$h<d$ describes {\it solitons} of dimension $d-h-1$; for $h=d$,
instead,
it describes {\it instantons} of the Euclidean version of the model.
These finite-energy (action) classical solutions are characterized by
radii
$r_i=1/M_i$, where $M_i$ are the various masses associated with
spontaneous
symmetry breakdown. From the point of view of the low-energy
effective theory
with symmetry group $H$, valid on energy
scales much smaller than $min\{ M_i \}$,
they can be viewed essentially as singularities of dimension $d-h-1$
in
$R^d$ (for solitons) or point singularities in $R^{d+1}$ (for
instantons) \raj.
The low-energy effective action is then well-defined only outside
these
singularities, which contain lumps of energy (action) involving
higher-lying
fields.

{}From now on we shall specialize to {\it stable topological
defects}, for which
the relevant homotopy groups $\Pi _h \left( G/H \right) =\IZ$. In
this case
there is an analytic topological invariant that can be written as
\eqn\ati{\int _{S_h} \ \omega_h \ ,}
where $S_h$ is an $h$-dimensional sphere surrounding the singularity
on
an $(h+1)$-dimensional hyperplane perpendicular to it and
$\omega_h $ is an $h$-form, which is exact: $\omega_h = d\phi_{h-1}
$, `outside' $S_h$.
Both $\phi_{h-1} $ and $\omega_h $ are constructed in terms of the
fundamental fields
of the low-energy effective theory with symmetry group $H$.

Consider now this low-energy effective theory in the presence of a
few
topological defects. Essentially we have a model defined on
$(d+1)$-dimensional
Minkowski space-time with a few $(d-h-1)$-dimensional holes in space
where the
topological defects live. In the case of instantons we would have
correspondingly a model on $(d+1)$-dimensional Euclidean space with a
few holes
at the location of the instantons. When the number of topological
defects
grows, the manifold on which the low-energy effective action is well
defined
 soon becomes very complicated. The question we would like to address
is what
happens if, for a certain range of parameters, the dynamics favours
the
formation of a macroscopic number of topological excitations with a
finite
density. The Julia--Toulouse theory provides a prescription to
identify the
additional hydrodynamical (long-lived) modes of a solid state medium
due to
the continuous distribution of topological defects. In the framework
of relativistic field theories these modes
would be additional fields in the new phase
of the low-energy theory at finite density of topological defects.

We start with the following observation. Associated with the integral
invariant \ati \ there is a $(d-h)$-form $J_{d-h}=\Omega_{h+1} ^*=
(d\omega_h )^*$. When
considered as forms defined only on the manifold with holes where the
low-energy
effective theory lives, both $\Omega_{h+1} $ and $J_{d-h}$ vanish
identically since $\omega_h
=d\phi_{h-1} $ there. If they are extended to the whole manifold
$R^{d+1}$, however,
they pick up delta-like singularities at the locations of the
topological
defects. In this case the totally antisymmetric components $J^{\mu
\nu
\dots \alpha }$ of $J_{d-h}$ describe the topologically conserved
``currents'' of
topological defects,
\eqn\tcctd{\partial _{\mu } J^{\mu \nu \dots \alpha } = 0.}
For point-like solitons $h=d-1$ and $J$ is a 1-form whose components
$J^{\mu }$
form a proper current. For string-like vortices $h= d-2$ and
$J_{d-h}$ is a 2-form
with components $J^{\mu \nu}$: the three pure space components
correspond to
the density of vortices, while the three mixed components correspond
to the
current density of vortices. For instantons $h=d$ and $J_{d-h}$ is a
0-form describing the density of instantons in Euclidean space:
in this case there is clearly no differential conservation law.

The idea of Julia and Toulouse to identify the additional low-energy
modes
in the phase with a finite density of topological defects consists in
promoting
the closed form $\omega_h$ defined on the very complicated manifold
`outside' the many defects to a fundamental form defined everywhere
on
$R^{d+1}$. In this way, the components of $J_{d-h}$ become smooth
fields on $R^{d+1}$ describing the conserved fluctuations of the
continuous
distribution of topological defects. These fluctuations constitute
the new
hydrodynamical modes.
Actually, since $J_{d-h}=(d\omega_h )^*$, these new
degrees of freedom are associated only with the {\it gauge-invariant}
part of the new fundamental $h$-form $\omega_h$.
For example, for vortices in (3+1) dimensions,
we  generically obtain 2 new degrees of freedom.

\newsec{Application to antisymmetric tensor field theories}
While the Julia--Toulouse prescription is simple and appealing for
the
identification of the additional low-energy modes due to a continuous
distribution of topological defects, it is not so simple to extract
the
dynamics of these new modes and their coupling to the original fields
of
the low-energy effective model. Two are the difficulties in the
framework
of ordered solid-state media. First, in the generic case, the form
$\omega_h $
is a complicated expression in terms of the fundamental fields
of the original theory; secondly, one has to introduce appropriate
dissipation terms for the new modes.

In this section we wish to point out that there is a class of
relativistic field theories for which the Julia--Toulouse
prescription can be made much
more precise: in these cases we will also obtain a simple form for
the effective action in the phase with finite density of topological
defects. These field theories are relevant to confinement physics,
the
low-energy limit of fundamental string theories and possibly
cosmology.

The class of field theories we have in mind contain in their
low-energy
limit a {\it compact} \poly\ {\it fundamental} $(h-1)$-form
$\phi_{h-1} $ with (generalized)
gauge invariance under transformations $\phi_{h-1} \to \phi_{h-1} +
d\lambda_{h-2} $.
In particular we will consider the following generic low-energy
effective
action:
\eqn\gleea{S=\int {(-1)^{h-1}\over e^2} \ d\phi_{h-1} \wedge
(d\phi_{h-1} )^* +
\kappa \ \phi_{h-1} \wedge j_{h-1}^* + S_M  \ ,}
where $j_{h-1}$ describes a conserved (tensor) current of fields
whose dynamics
is governed by the action $S_M$. For convenience we will take the
canonical
mass dimension of $\phi_{h-1} $ as $(d-1)/2$, so that $e^2$ is a
dimensionless
coupling constant. Gauge fixing is always understood.
For $j_{h-1}=0$ this action describes
\eqn\pdf{N_{\phi_{h-1}}= \left( {d\atop h-1} \right)
-\left( {d\atop h-2}\right)
+\left( {d\atop h-3} \right)
+\dots + (-1)^{h-1} \left( {d\atop 0} \right) = \left( {d-1\atop h-1}
\right) }
{\it massless} physical degrees of freedom.

The compactness \poly\ of the form $\phi_{h-1} $ ensures the presence
of $(d-h-1)$-dimensional singularities describing the topological
defects.
The origin of these can be thought of as spontaneous symmetry
breaking at
a very high energy scale or as a different mechanism. For example,
the
field theories describing the low-energy limit of fundamental string
theories
typically contain higher-rank tensor fields \gsw: in this case the
topological
defects of the low-energy field theory may be thought of as lumps of
energy (action) involving higher-lying string modes and describe
essentially
$(d-h-1)$-branes \dkl. For our purposes, however, the origin of the
topological
defects is inessential; the important point is that in this case
the topological invariant \ati \ can be formulated directly in terms
of the form $\omega_h = d\phi_{h-1} $, which is {\it linear} in the
fundamental
field $\phi_{h-1} $ of the low-energy effective field theory.
Moreover, in the
framework of relativistic field theories one clearly does not have
the problem
of introducing dissipation terms.

In the following we shall follow closely the idea of Julia and
Toulouse, i.e.
we shall consider that a condensation of topological defects
generates a new
low-energy mode described by the gauge-invariant part of an $h$-form
$\omega_h $: however, we will take at first $\omega_h $ to be a new,
additional
field, which is not related to $d\phi_{h-1} $. Therefore the action
in the phase
with finite density of topological defects will depend on this
additional
field $\omega_h $, as well as on the original low-energy fields
$\phi_{h-1} $.

In order to write down this action, we start by noting that the
condensation of topological defects generates
a new scale $\Lambda $ in the theory.
Suppose that solitons of dimension $d-h-1$ condense. Take any
$(h+1)$-dimensional hyperplane in $R^d$: generically, the
intersections
of this hyperplane with the solitons will be point-like. The new
parameter can
be taken to represent the average density $\rho $ of these points on
the
chosen hyperplane. Since $\rho $ has dimension $({\rm mass})^{h+1}$,
the new
mass scale $\Lambda $ is essentially given by $\Lambda \propto \rho
^{1/h+1}$.
This formula is valid also in the case of instantons, for which $h=d$
and
$\rho $ describes the density of instantons in Euclidean space.

With this point of view there are three requirements on the effective
action in the phase at finite density of topological defects.
The first is {\it gauge invariance}. Actually two gauge symmetries
must
be present: the first is the original gauge symmetry under
transformations
$\phi_{h-1} \to \phi_{h-1} +d\lambda_{h-2} $; the second is a new
gauge symmetry under
transformations $\omega_h \to \omega_h +d\psi_{h-1} $, which ensures
that only the
gauge invariant part of $\omega_h $ describes a new physical degree
of freedom.
The second requirement is {\it relativistic invariance}. The
third is that in the limit $\Lambda \to 0$ one has to recover the
original
low-energy theory. Up to two derivatives in the new fundamental form
$\omega_h $
we  thus obtain the effective action
\eqn\eafd{\eqalign{S_{d-h-1}&=\int {(-1)^h\over \Lambda ^2}
\Omega_{h+1} \wedge \Omega_{h+1} ^*
+{(-1)^{h-1}\over e^2} (\omega_h -d\phi_{h-1}) \wedge (\omega_h
-d\phi_{h-1} )^*
\cr
& +\kappa (\omega_h -d\phi_{h-1} )\wedge T_h^* +S_M \ ,\cr }}
where we index the new action by the dimension of the condensing
topological defects and
$\Omega_{h+1} \equiv d\omega_h $. Relativistic invariance and the two
gauge
symmetries are manifest. Actually, transformations $\omega_h \to
\omega_h +
d\psi_{h-1} $ must be accompanied by corresponding transformations
$\phi_{h-1}
\to \phi_{h-1} + \psi_{h-1}$, so that the full new gauge symmetry is
given by
\eqn\ngs{\eqalign{\omega_h &\to \omega_h +d\psi_{h-1} \ ,\cr
\phi_{h-1} &\to \phi_{h-1} +\psi_{h-1} \ .\cr }}
Correspondingly, the original conserved $(h-1)$-form current must be
promoted to an $h$-form $T_h$ such that
\eqn\nmf{\partial_{\mu }T^{\mu \nu \dots \alpha }= {1\over h} \
j^{\nu \dots
\alpha } \ .}
In the limit $\Lambda \to 0$ the only contributions to the partition
function come from configurations for which $\Omega_{h+1} =0$. The
solution to
this constraint is $\omega_h =d\psi_{h-1}$: the field $\psi_{h-1} $
can then be absorbed
into $\phi_{h-1} $ and one recovers (upon an integration by parts)
the original
low-energy effective action \gleea .

Let us now have a closer look at the action \eafd . Clearly the new
gauge
symmetry \ngs \ has to be gauge fixed, i.e. one has to divide
the partition function corresponding to \eafd \ by the gauge volume.
As usual for Abelian systems, this
means that the functional integration over the original
field $\phi_{h-1} $ can be dropped after having absorbed $d\phi_{h-1}
$ into a redefinition of
the new field $\omega_h $. The gauge-fixed action can thus be
formulated
exclusively in terms of the new field $\omega_h $:
\eqn\gffd{S_{d-h-1}=\int {(-1)^h\over \Lambda^2} \ \Omega_{h+1}
\wedge \Omega_{h+1} ^*
+{(-1)^{h-1}\over e^2} \ \omega_h \wedge \omega_h ^* + \kappa \
\omega_h
\wedge T_h^* +S_M \ .}
For $T_h=j_{h-1}=0$ the equations of motion following from this
action are
given by
\eqn\neom{\eqalign{\partial _{\mu }\partial ^{[\mu} \omega^{\alpha _1
\dots \alpha _h ]} &+ {\Lambda ^2\over e^2} \ \omega ^{\alpha _1
\dots
\alpha _h} = 0\ ,\cr
\partial _{[\mu } \omega _{\alpha _1\dots \alpha _h ]} &\equiv
\partial _{\mu }\omega _{\alpha _1 \dots \alpha _h} +(-1)^h
\partial _{\alpha _1} \omega _{\alpha _2 \dots \alpha _h \mu } \cr
&+\partial _{\alpha _2}\omega _{\alpha _3 \dots \mu \alpha _1}
+\dots +(-1)^h \partial _{\alpha _h} \omega_{\mu \alpha _1 \dots
\alpha _{h-1}} \ .\cr }}
Taking derivatives with respect to all variables $x^{\alpha _i}$ we
then
obtain
\eqn\con{\partial _{\alpha _i} \omega ^{\alpha _1\dots \alpha _h} =0\
,
\qquad \qquad \forall i\ .}
Inserting this back into \neom \ we finally get
\eqn\ffeq{\eqalign{ \left( \partial ^2 + m^2 \right) \ &\omega
^{\alpha _1
\dots \alpha _h} = 0\ ,\cr
m &= {\Lambda \over e} \ ,\cr }}
showing explicitly that the physical content of \gffd \ consists of
\eqn\odf{N_{\omega_h }= \left( {d\atop h} \right) }
{\it massive} degrees of freedom \foot{Massive antisymmetric tensors
have been previously studied in \refs{\topmass,\att,\bdqq}.}.

What has happened here is the exact contrary of the familiar Higgs
mechanism, where the original gauge field `eats' the Goldstone mode
due to the condensation of the Higgs field thereby acquiring a
longitudinal
part and a mass. Here it is the new gauge field, due to the
condensation
of topological defects, which `eats' the original gauge field,
thereby
acquiring $\left( {d-1\atop h}\right)$ `longitudinal' degrees of
freedom and a mass.
This renders much more precise the original Julia--Toulouse
prescription
of `promoting the $(h-1)$-form $\phi_{h-1} $ to a new fundamental
$h$-form
$\omega_h $'. Actually, the
relation between this mechanism, which we shall call the {\it
Julia--Toulouse
mechanism} and the Higgs mechanism can be made more precise. Indeed,
in the
next section we will show that the Julia--Toulouse mechanism is the
exact {\it dual} to the Higgs mechanism.

 The question of additional couplings  remains to be discussed.
As we have shown above, the original, conserved $(h-1)$-form current
$j_{h-1}$
must be promoted to an $h$-form $T_h$ satisfying \nmf .
 By integrating out
the field $\omega _h$ we obtain an induced action for $T_h$. The
non-local
kernel in this action is short-range since $\omega _h$ is a massive
field.
By taking the local limit we obtain  a local, low--energy, induced
action $S_{\rm ind}\left( T_h \right) $ for $T_h$.
%as we shall illustrate below in
%a concrete example, this action describes $\left( {d\atop
% h-1}\right) $
%massive degrees of freedom.
This example will  clarify that the
Julia--Toulouse mechanism describes a confinement phase for the
original $(h-1)$-form $\phi _{h-1}$.

%The original matter action $S_M$ can also be
%expressed in terms of the field $T_h^*$ by introducing a Lagrange
% multiplier
%$\lambda _{h-1}$ in the partition function as follows:
%\eqn\lmpf{\int {\cal D}\lambda _{h-1}\ {\cal D}T_h^*
%\ {\cal D}({\rm matter \ fields})
%\ {\rm exp}\left\{ iS_M+i\int \lambda _{h-1}
%\wedge \left( j_{h-1}^* -dT_h^* \right)
%\right\} \ .}
%Integrating out the matter fields and the Lagrange multiplier we
% obtain
%an expression for $S_M\left( T_h^* \right) $ which provides
% corrections to
%$S_{\rm ind}\left( T_h^* \right) $. Naturally this procedure can be
% carried %out exactly only if the original $S_M$ is quadratic in the
% current $j$; %otherwise
%we have to rely on perturbation theory.

We still have to discuss  the possible
{\it topological terms} that we have neglected until now. Depending
on
the space-time dimensionality, there are three possible types of
gauge invariant topological terms:
\eqn\tote{\eqalign{d &=2h-2\ , \qquad \qquad \qquad \phi_{h-1} \wedge
d\phi_{h-1} \ ,\cr
d &= 2h-1\ , \qquad \qquad \left( \omega_h -d\phi_{h-1} \right)
\wedge \left(
\omega_h -d\phi_{h-1} \right) \ ,\cr
d &=2h\ , \qquad \qquad \qquad \qquad \omega_h \wedge d\omega_h \
.\cr }}
The first is a (generalized) Chern--Simons term for the original
gauge field
$\phi_{h-1} $. We do not include such a term in our generic action
\gleea \ since
it produces a confinement mechanism for the topological defects which
prevents a condensation phase. This is well known \pis\ for the
familiar vector
Chern--Simons term in $d=2$, which suppresses the instantons of the
model.
The same type of argument, however, leads to similar results in
higher-dimensional theories. The second gauge-invariant topological
term
comprises actually three terms: a topological mass term $\omega_h
\wedge \omega_h$
for $\omega_h $, a so-called BF-coupling $\omega_h \wedge
d\phi_{h-1}$, and a
(generalized) $\theta $-term $d\phi_{h-1} \wedge d\phi_{h-1} $ for
$\phi_{h-1} $. According
to our construction this gauge-invariant combination  appears in the
action at finite density of topological defects if the $\theta$-term
$d\phi_{h-1} \wedge d\phi_{h-1} $ is present in the original
low-energy effective action.
We will show below, how this affects the mass \ffeq \ in a specific
example.
The third possible topological term is a (generalized) Chern-Simons
term
for $\omega_h $. This term cannot be excluded on general grounds: it
could appear
in the actions \eafd \ and \gffd \ if the condensation of topological
defects
violates some discrete symmetries of the original theory or if these
were
anyhow violated by the coupling with $j$.

We would like to conclude this section by stressing that
the realization of a phase with a condensate of topological
defects remains a dynamical issue, which cannot be solved within the
present
framework. Here we have  only discussed the form of the field
theories at
finite density of topological defects, assuming that a condensation
indeed takes place.
The best way to address the condensation mechanism is via a lattice
formulation of the antisymmetric tensor theories
\refs{\kt,\eli,\bmk,\poly,\cardy,\orland,\savit}\ although
a full renormalization group analysis is available only in 2D \kt.

\newsec{Compact QED in (3+1) dimensions}
Before pursuing our general analysis we shall pause to describe a
first
concrete example, namely {\it compact} quantum electrodynamics
(QED)\refs{\bmk,\poly}.

Compact QED (in 3+1 dimensions) is QED with {\it magnetic monopoles}
\godol.
 It can
be thought of as a cut-off theory describing the low-energy sector of
an
$SO(3)$ Georgi--Glashow model with  spontaneously broken symmetry
$SO(3)\to U(1)$. From the point of view of the low-energy $U(1)$
theory
the 't Hooft--Polyakov monopoles of the model appear essentially as
singular
Dirac monopoles labelled by $\Pi _2 \left( SO(3)/U(1) \right) =
\Pi _1 \left( U(1) \right) = \IZ$.

It is by now generally accepted that the condensation of monopoles in
compact QED drives a transition from a weak coupling Coulomb phase to
a
strong coupling {\it confinement phase}, characterized by a `massive
photon'
and an area law for the Wilson loop \poly. An analytical proof of
this mechanism has
been recently given for $N=2$ supersymmetric Yang-Mills theories \sw,
for which
the role of the Higgs field is played by the scalar in the $N=2$
vector
multiplet.

In the Coulomb phase, magnetic monopoles are dilute. Away from the
singularities
the action is given simply by
\eqn\acqed{\eqalign{S &=\int d^4x \ {-1\over 4e^2} F_{\mu \nu }F^{\mu
\nu }
+{\theta \over 32 \pi ^2} F_{\mu \nu } {F^{\mu \nu }}^* \ , \cr
F_{\mu \nu } &\equiv \partial _{\mu }A_{\nu } - \partial_{\nu }
A_{\mu } \ ,
\qquad \qquad {F^{\mu \nu }}^* \equiv {1\over 2} \epsilon ^{\mu \nu
\alpha
\beta } F_{\alpha \beta } \ ,\cr }}
where we have included also a $\theta $-term to illustrate how this
affects
the Julia--Toulouse mechanism. The $\theta $-term can be rewritten as
a total
derivative: in the absence of magnetic monopoles it can be dropped
altogether;
in their presence, however, it produces non-trivial effects,
notably it assigns an electric charge $q= e\theta /2\pi $ to
elementary
magnetic monopoles \ewit; $\theta $ is an angular variable:
elementary monopoles
with $\theta = 2\pi n + \theta '$ are equivalent to dyons with
electric charge
$en$ and parameter $\theta '$. Since we are interested here in the
condensation
of magnetic monopoles, we shall restrict $\theta $ to the range
$0\le \theta <2\pi $.

Following the general construction outlined in the preceding sections
we can
now immediately write down the action for compact QED in the phase
with a
monopole condensate. The gauge-fixed action is formulated in terms of
an
antisymmetric tensor $\omega _{\mu \nu }$ and reads
\eqn\qedfdm{\eqalign{S_{1} &= \int d^4x \ {1\over 12 \Lambda ^2}
\Omega _{\mu \nu \alpha}
\Omega ^{\mu \nu \alpha } - {1\over 4e^2} \omega _{\mu \nu }\omega
^{\mu \nu }
+{\theta \over 64 \pi ^2} \omega _{\mu \nu }\epsilon ^{\mu \nu \alpha
\beta}
\omega _{\alpha \beta} \ ,\cr
\Omega _{\mu \nu \alpha} &\equiv \partial_{\mu }\omega _{\nu \alpha }
+
\partial _{\nu } \omega _{\alpha \mu } +\partial _{\alpha }\omega
_{\mu \nu }
\ .\cr }}
The equations of motion following from this action are given by
\eqn\eofqed{\partial_{\mu } \Omega ^{\mu \alpha \beta} + {\Lambda
^2\over e^2}
\omega ^{\alpha \beta} - {\Lambda ^2\theta \over 16 \pi ^2} \epsilon
^{\alpha
\beta \gamma \delta } \omega _{\gamma \delta} =0 \ .}
Actually, only the three equations for the space--space components
$\omega ^{ij}$
are true equations of motion. The remaining three equations are
constraints
enforced by the three Lagrange multipliers $\omega ^{0i}$.

Contracting \eofqed \ with $\partial _{\alpha}$ we obtain the
conditions
\eqn\qedcon{\eqalign{\partial _{\mu }\omega ^{\mu \nu } &+ {e^2
\theta \over
8\pi ^2} \Omega ^{\nu} =0 \ ,\cr
\Omega ^{\mu } &\equiv {1\over 6} \epsilon ^{\mu \nu \alpha \beta }
\Omega _{\nu \alpha \beta } \ .\cr }}
Contracting then the equations of motion
 \eofqed \ with $\epsilon _{\nu \gamma
\alpha \beta} \partial ^{\gamma }$ and using the above conditions we
finally obtain
\eqn\ffeof{\eqalign{&\left( \partial ^2 + m_{\theta}^2 \right)
\Omega ^{\mu } =0 \ ,\cr
&m_{\theta }= {e\Lambda \over 4\pi} \
\sqrt{ \left( {4\pi \over e^2} \right) ^2 + \left(
{\theta \over 2\pi }\right) ^2 } \ .\cr}}
As expected $S_{1}$ describes a {\it massive vector particle}
(`massive photon'). Note that the mass of this particle is determined
by the same modular parameter $\tau = (\theta /2\pi ) + i(4\pi
/e^2)$, which enters the mass
formula for the monopoles in the BPS limit \olive.

In the following we shall consider matter couplings. To this end we
add a coupling term
\eqn\cot{\eqalign{S^E_{\rm coup} &= i \int d^4x \ \omega _{\mu \nu }
T_{\mu \nu }
\ ,\cr
\partial _{\nu } T_{\mu \nu } &= {1\over 2} j_{\mu } \ ,\cr }}
to the Euclidean version of the action \qedfdm \ and we integrate out
$\omega _{\mu \nu }$ to obtain an induced action for $T_{\mu \nu }$.
Note
that for $\Lambda \to 0$ we have $\omega _{\mu \nu } \to \partial
_{\mu }
A_{\nu } - \partial _{\nu } A_{\mu }$ and therefore \cot \ reduces to
the standard photon coupling to the conserved matter current $j_{\mu
}$.

The result of the Gaussian integration is the Euclidean induced
action
\eqn\eia{S^E_{\rm ind}=\int d^4x \ \Lambda ^2 T_{\mu \nu} {1\over
m_{\theta}^2-
\nabla ^2} T_{\mu \nu } + 2e^2 \partial _{\nu }
T_{\mu \nu } {1\over m_{\theta }^2-
\nabla ^2} \partial _{\alpha }T_{\mu \alpha } + i{e^2\Lambda ^2\theta
\over
16\pi ^2} T_{\mu \nu } {\epsilon _{\mu \nu \alpha \beta} \over
m_{\theta} ^2 -
\nabla ^2} T_{\alpha \beta }\ .}
On distance scales much smaller than $1/\Lambda $ we can neglect
terms
$\Lambda ^2/\nabla ^2$ and the induced action reduces to
\eqn\coulim{S^E_{\rm ind} = \int d^4x \ {e^2\over 2} \ j_{\mu }
{1\over -\nabla^2}
j_{\mu } \ ,}
which shows that at short distances we have the usual Coulomb
interactions
between charges. In the following, however, we shall be mainly
interested
in the opposite limit, the long-distance (low-energy) limit.

In order to establish the induced action in this limit we first note
that monopoles are expected to condense at strong coupling. In the
following we shall assume $e\gg 1$, so that we have  well separated
scales $e/m_\theta$, $1/m_\theta$, and $1/em_\theta$.
Moreover, we are assuming that $n\equiv \theta/2\pi\gg 4\pi/e^2$.
 We are seeking then a low-energy
induced action which is valid for scales $R\gg 1/em_\theta $ but
including
scales $R=O(e/m_\theta )$, $O(1/m_\theta)$. Technically this means
that in \eia \ we can
neglect terms $\nabla ^2/(em_\theta) ^2$ but we have to keep terms
$e^2\nabla ^2 /m_\theta ^2$ and $\nabla ^2 /m_\theta ^2$. In this
{\it local limit} the induced
action reduces to
\eqn\llia{S^E_{\rm ind}= \int d^4x \ {\Lambda ^2\over m_{\theta }^2}
\ T_{\mu \nu }
T_{\mu \nu } + {2e^2\over m_{\theta }^2} \ \partial _{\nu }
T_{\mu \nu } \partial _{\alpha }
T_{\mu \alpha } + i{e^2\Lambda ^2 \theta \over 16 \pi ^2 m_{\theta
}^2}
\ T_{\mu \nu }
\epsilon _{\mu \nu \alpha \beta}\left(1+\nabla ^2 /m_\theta ^2\right
) T_{\alpha \beta} \ .}

\subsec{Point particles}
In order to expose the nature of the monopole condensate phase we
first
consider external point particles and we compute the induced static
potential.
In this case we have
\eqn\cpp{j_{\mu } = \int d\tau \ {dx_{\mu }\over d\tau } \ \delta ^4
\left( {\bf x} - {\bf x}(\tau )\right) \ ,}
where ${\bf x}(\tau )$ parametrizes a closed curve in the
four-dimensional
Euclidean space. Correspondingly we have
\eqn\nvpp{\eqalign{T_{\mu \nu } &= {-1\over 2}
\int d^2 {\bf \sigma} \ X_{\mu \nu }
({\bf \sigma }) \ \delta ^4 \left( {\bf x}-{\bf x}({\bf
\sigma})\right) \ ,\cr
X_{\mu \nu } &= \epsilon ^{ab} \ {\partial x_{\mu } \over \partial
\sigma ^a}
{\partial x_{\nu } \over \partial \sigma ^b} \ ,\cr }}
where ${\bf x}({\bf \sigma })$ parametrizes a surface bounded by the
closed
curve ${\bf x}(\tau )$. The induced metric $g_{ab}$ on this surface
and its
determinant $g$ are given by
\eqn\inme{\eqalign{g_{ab} &= {\partial x_{\mu }\over \partial \sigma
^a}
{\partial x_{\mu }\over \partial \sigma ^b}\ ,\cr
g &= {1\over 2} X_{\mu \nu } X_{\mu \nu } \ .\cr }}

Inserting \nvpp \ into \eia \ and carefully taking the local limit
as explained above yields
\eqn\fiapp{S^E_{\rm ind} = {\Lambda ^2 K_0(\epsilon m_\theta ) \over
4\pi }
\int d^2 {\bf \sigma } \ \sqrt{g} + {e^2 m_\theta f(\epsilon m_\theta
)\over 8\pi ^2  }
\int d\tau \ \sqrt{{dx_{\mu }\over d\tau } {dx_{\mu }\over d\tau }
}-i{\pi\over n}\ \nu \ ,}
where $f(x)\equiv \int_x^\infty {dz\over z}K_1(z)$ and $K_0$ and
$K_1$ are
modified Bessel functions.
Here $\epsilon $ is a short-distance (ultraviolet) cutoff satisfying
$ \epsilon m_\theta \ge O(1) $ and
\eqn\inters{\nu={1\over 4\pi}\int d^2\sigma\ \sqrt{g}\
\epsilon^{\mu\nu\alpha\beta} g^{ab}\partial_a t_{\mu\nu}
\partial_b t_{\alpha\beta}\ ,}
is the self-intersection number of the surface $x(\sigma)$ in
four Euclidean dimensions, defined in terms of  $t_{\mu\nu}\equiv
X_{\mu\nu}/\sqrt{g}$.
%It is needed, in the case of point particles,
%in an effective action valid only on scales larger than $1/\Lambda
% $.

The first term in the induced action \fiapp \ is the Nambu -Goto term
measuring the {\it area} of the surface enclosed by ${\bf x}(\tau )$.
It
shows that at large distances the potential between a
particle-antiparticle pair is linear and identifies the monopole
condensate
phase as a {\it confinement phase}. The second term is a correction
term
measuring the length of the boundary of the surface. %Although it is
% negligible
%on scales $R\gg e/\Lambda $, its negative sign indicates the
% presence of a
%minimum of $S^E_{\rm ind}$ on a scale $R=O\left( e/\Lambda \right)
% $.
The third is the `spin' term \poly. Note that the string
becomes `fermionic' for $\theta/2\pi=n=1$ which corresponds to the
condensation
of elementary dyons.
Let us also mention that the above computation can be generalized.
Had we started
with an $(h-1)$-form coupled to a closed $(h-1)$-dimensional
hypersurface in
$(d+1)$-dimensional Euclidean space, the dominant piece of $S^E_{\rm
ind}$
at long distances in the phase with a condensate of topological
defects
would have been the $h$-dimensional hypervolume enclosed by the
$(h-1)$-dimensional closed hypersurface. This shows that the
Julia--Toulouse
mechanism for the phase with a condensate of topological defects
describes the
natural generalization of a {\it confinement phase}.

\newsec{The Julia--Toulouse mechanism as the dual Higgs mechanism}

\subsec{Standard Duality}

It is by now well known that in $(d+1)$-dimensional space-time an
$(h-1)$-form
$\phi_{h-1} $ is {\it dual} to a $(d-h)$-form $\tilde\phi_{d-h}$.
Indeed, starting from the master action \foot{Throughout this section
we will consider only the gauge sector, leaving out couplings to
conserved
currents.}
\eqn\maaphi{S_{\rm master}= \int {(-1)^{h-1}\over e^2} \ \Phi_h
\wedge \Phi_h ^*
+\Phi_h \wedge d\tilde\phi_{d-h} \ ,}
our action \gleea \ can be recovered
 by integrating out the Lagrange multiplier
$\dphi$ and solving the resulting constraint as $\Phi_{h}
=d\phi_{h-1} $.
On the other hand we can also integrate out $\Phi_h $ to obtain an
action
for the dual form $\dphi$:
\eqn\daphi{\tilde S = \int {(-1)^{d-h} e^2\over 4}\ d\dphi \wedge
\left( d\dphi
\right) \ .}
This action describes
\eqn\ddf{N_{\dphi} = \left( {d\atop d-h} \right) - \left( {d\atop
d-h-1}
\right) +\left( {d\atop d-h-2}\right) +\dots +(-1)^{d-h}\left(
{d\atop 0}
\right) = \left( {d-1\atop d-h} \right) = N_{\phi_{h-1} } }
massless degrees of freedom. The two actions
$S$ and $\tilde S$ are related by
a functional Fourier transform and constitute two different
representations
of the same physics.

\subsec{Duality including topological defects}

The duality transformation can be extended to {\it compact}
antisymmetric tensor theories. In particular, we shall consider both
the gauge field $\phi _{h-1}$ and its dual $\tilde \phi _{d-h}$ to be
compact, so that the model admits two types of topological defects:
in
modern parlance the original $(d-h-1)$-branes and their dual
$(h-2)$-branes
\dkl. Some particular examples of duality in presence of topological
defects have already been considered in \klee.

The best way to formulate the extended duality is to treat these
topological defects explicitly. We  thus start from the master action
\eqn\exdu{S_{\rm master}= \int {(-1)^{h-1} \over e^2} \left(\Phi _h
-q\, V_h \right) \wedge \left( \Phi _h - q\, V_h \right) ^* +
\Phi _h \wedge \left( d\tilde \phi _{d-h} - \tilde{q}\, \tilde
V_{d+1-h} \right)
 \ ,}
where $V_h$ and $\tilde V_{d+1-h}$ are singular $h$-
and $(d+1-h)$-forms such that
\eqn\ccbra{\eqalign{j_{d-h} &= (-1)^{d+h^2} \ \left( dV_h \right) ^*\
,\cr
\tilde j_{h-1} &=  (-1)^{(d+1)(d-h)} \
\left( d\tilde V_{d+1-h} \right) ^* \ ,\cr }}
represent the conserved `currents' of the $(d-h-1)$-branes and their
dual $(h-2)$-branes.
Here $q$ and $\tilde q$ are constants of canonical
dimension $\pm (d-2h+1)/2$ respectively; they play the role of the
units of charge for the topological defects and their duals. A useful
representation is given in terms of $V^*$ and $\tilde{V}^*$ as
follows:
\eqn\vv{\eqalign{
V_h^*= T_{d-h+1} \ , \cr
\tilde{V}_{d-h+1}^*={-(-1)^{h(d+1-h)}}\, \tilde{T}_h\ ,\cr}}
where
\eqn\tt{\eqalign{
%% FOLLOWING LINE CANNOT BE BROKEN BEFORE 70 CHAR
T_{d-h+1}^{\mu_1\cdots\mu_{d-h+1}}&=\int\delta^{d+1}\left(x-y(\sigma)\right)
\, dy^{\mu_1}\wedge\cdots\wedge y^{\mu_{d-h+1}}\ ,\cr
%% FOLLOWING LINE CANNOT BE BROKEN BEFORE 70 CHAR
\tilde{T}_h^{\mu_1\cdots\mu_{h}}&=\int\delta^{d+1}\left(x-\tilde{y}(\tilde\sigma 
)\right)
\, d\tilde{y}^{\mu_1}\wedge\cdots\wedge \tilde{y}^{\mu_{h}}\ ,}}
and $y(\sigma)$ and $\tilde{y}(\tilde\sigma)$ represent
open hypersurfaces bounded by the world-hyperlines of the topological
defects and their duals.

By integrating out $\tilde \phi _{d-h}$ in \exdu\  we obtain the
action
\eqn\extode{S= \int {(-1)^{h-1}\over e^2} \left( d\phi _{h-1} - q\,
V_h
\right) \wedge \left( d\phi _{h-1} - q\, V_h \right) ^* + \tilde
q\phi _{h-1} \wedge
\tilde j_{h-1}^*  \ .}
The singular $V_h$ can be absorbed into $d\phi _{h-1}$ by considering
it
as a singularity due to the compactness of the gauge field and
represents
a solitonic $(d-h-1)$-brane. The dual $(h-2)$-branes, instead, appear
as
elementary non-dynamical matter and couple `minimally' to $\phi
_{h-1}$.

By integrating out $\Phi _h$, instead, we obtain the dual action
\eqn\duextode{\eqalign{\tilde S &= \int {(-1)^{d-h}e^2\over 4} \left(
d\tilde \phi _{d-h} - \tilde q\, \tilde V_{d+1-h} \right) \wedge
\left(
d\tilde \phi _{d-h} - \tilde q\, \tilde V_{d+1-h} \right) ^* \cr
&+ q\, \tilde \phi _{d-h}
\wedge j_{d-h} ^* - q\tilde q\,  V_h \wedge \tilde V_{d+1-h} \ .\cr}}
In this formulation of the theory it is the $(h-2)$-branes,
represented by
$\tilde B_{d+1-h}$, which appear as topological defects, while the
original
$(d-h-1)$-branes enter as elementary non-dynamical matter `minimally'
coupled to $\tilde \phi _{d-h}$.
The last term in \duextode\ represents a generalized Aharonov--Bohm
interaction
between the topological defects. Using the representations \vv, \tt,
it is easy to see that it contributes to the partition function a
term:
\eqn\dirac{\exp\left\{i(-1)^{d+1}\, q\, \tilde q\, I(y,\tilde
y)\right\}
\ ,}
where $I$ is the signed intersection number of the two hypersurfaces
$y$ and $\tilde y$ in ($d+1$)-dimensional space-time. Requiring that
this term does not contribute to the partition function leads to a
generalized Dirac quantization condition
\claudiot:
\eqn\quant{q\, \tilde q = 2\pi p , \qquad p\in \IZ.}
The above exact dualities are generically broken in the
presence of
dynamical matter.

\subsec{Duality of Higgs and confinement phases}

In the following we shall perform an analogous duality transformation
on the
finite density action $S_{d-h-1}$ in \eafd .

To this end we start again from a master action,
\eqn\mafd{\eqalign{S_{\rm master}&= \int {(-1)^h\over \Lambda ^2}
\Omega_{h+1} \wedge
\Omega_{h+1} ^* + \Omega_{h+1} \wedge \left( d\omegad - \dphi\right)
\cr
&+{(-1)^{h-1}\over e^2} (\omega_h -d\phi_{h-1} )\wedge (\omega_h
-d\phi_{h-1} )^* -
\omega_h \wedge d\dphi \ ,\cr}}
formulated in terms of the dual couples $\phi_{h-1} $, $\dphi$ and
$\omega_h $,
$\omegad$ and the additional master field $\Omega_{h+1} $.
This action possesses two gauge symmetries
under transformations
\eqn\fgsy{\eqalign{\omega_h &\to \omega_h +d\psi_{h-1} \ ,\cr
\phi_{h-1} &\to \phi_{h-1} + \psi_{h-1} \ ,\cr }}
and transformation
\eqn\sgsy{\eqalign{\omegad &\to \omegad + \tilde\psi_{d-h-1}\ ,\cr
\dphi &\to \dphi + d\tilde\psi_{d-h-1 } \ .\cr }}
Clearly, these two gauge symmetries are also dual to each other. Both
have
to be gauge-fixed.

Let us first integrate out the fields $\dphi$, $\omegad$
and $\Omega_{h+1} $. To this end
we first absorb $d\omegad$ into a redefinition of $\dphi$: the
remaining
divergent integration over $\omegad$ is cancelled out by the gauge
fixing.
The integration over the Lagrange multiplier $\dphi$ enforces
the constraint
\eqn\rcos{\Omega_{h+1} = (-1)^h d\omega_h \ .}
The last integration over $\Omega_{h+1} $  then yields
\eqn\eafda{S_{d-h-1}=\int {(-1)^h\over \Lambda ^2} \Omega_{h+1}
\wedge \Omega_{h+1} ^*
+{(-1)^{h-1}\over e^2} (\omega_h -d\phi_{h-1}) \wedge (\omega_h
-d\phi_{h-1} )^* \ ,}
with $\Omega_{h+1}$ as in \rcos, which is exactly the action
\eafd \ for the confinement phase at finite density of topological
defects in the case $T_h=0$.

On the other hand, we can integrate out the fields $\phi_{h-1} $,
$\omega_h$ and
$\Omega_{h+1} $ to obtain an action for the dual variables. The
integration over
$\phi_{h-1}$ is eliminated by gauge-fixing after absorbing
$d\phi_{h-1} $ into a
redefinition of $\omega_h$. The two remaining integrations over
$\omega_h$ and
$\Omega_{h+1}$ then give the action
\eqn\dafd{\eqalign{\tilde {S}_{d-h-1}&= \int{(-1)^{d-h} e^2\over 4} \
d\dphi \wedge \left(
d\dphi \right) ^* \cr
&+{(-1)^{d-h-1} \Lambda ^2\over 4} \left( \dphi -
d\omegad\right) \wedge \left( \dphi-d\omegad \right) ^* \ ,\cr }}
where gauge fixing of the dual gauge symmetry is always understood.

For the particular case $h=d-1$,  $\dphi$ is a 1-form and $\omegad$
is
a 0-form: in this case ${\tilde S_{0}}$ embodies the familiar Higgs
mechanism,
in which a vector gauge field `eats' a scalar thereby acquiring a
mass,
in
its pristine St\"uckelberg formulation \stu. As
a natural generalization we shall call the theory
described by ${\tilde S_{d-h-1}}$
the Higgs phase for a $(d-h)$-form $\dphi$. As shown by the above
computation
this (generalized) Higgs mechanism is the dual of the Julia--Toulouse
mechanism, embodied by $S_{d-h-1}$ and describing the confinement
phase for the dual $(h-1)$-form $\phi_{h-1} $. In this duality
transformation all coupling
constants are reversed: in particular, for $\Lambda \to 0$ we recover
the
familiar duality between $(h-1)$-forms and $(d-h)$-forms described at
the beginning of the section.
Thus we reach the conclusion that the same physics can be described
as a
confinement phase for $\phi_{h-1} $ or a Higgs phase for $\dphi$:
this generalizes
the well known 't Hooft--Mandelstam duality \man\ to
any compact $p$-form theory in $(d+1)$
space-time dimensions.

 So far we have two pairs of dual actions,
namely the pair \gleea\ --\daphi\ describing the `Coulomb phase' and
the
pair \eafda\ --\dafd\ describing the confinement phase for $\phi$ and
the Higgs phase for $\dphi$. We may wonder if there is a dual  phase
which would be a Higgs phase for $\phi$ and a confinement phase for
$\dphi$.
Clearly we expect such a phase to be generated by the condensation of
the dual $(h-2)$-branes. The best way to see this is to apply the
Julia--Toulouse mechanism to the dual action \daphi . To this end we
can
just repeat the steps leading to
equations \eafda\ --\dafd\
with the exchanges:
\eqn\exch{\eqalign{\phi_{h-1} &\leftrightarrow\dphi \
,\qquad\qquad\qquad
h  \leftrightarrow d-h+1 \ ,\cr  e^2  &\leftrightarrow 4/e^2\
,\qquad\qquad  \qquad\Lambda^2  \leftrightarrow
{\tilde\Lambda}^2\ .}}
 We  arrive at the two dual actions:
\eqn\deafda{\eqalign{S_{h-2}&=\int
{(-1)^{d-h+1}\over{\tilde\Lambda}^2}
d\rho_{d-h+1}\wedge \left( d\rho_{d-h+1}\right) ^*\cr
&+{(-1)^{d-h} e^2\over 4}\left(\rho_{d-h+1}-d\dphi\right)
\wedge\left(\rho_{d-h+1}-d\dphi\right)^*\ ,\cr
\tilde{S}_{h-2}&=\int {(-1)^{h-1}\over e^2}
d\phi_{h-1}\wedge \left( d\phi_{h-1}\right) ^*+
{(-1)^{h-2} {\tilde\Lambda}^2\over 4}\,
%% FOLLOWING LINE CANNOT BE BROKEN BEFORE 70 CHAR
\left(\phi_{h-1}-d\tilde\rho_{h-2}\right)\wedge\left(\phi_{h-1}-d\tilde\rho_{h-2 
}\right)^*.}}
Both  describe $N_{\rho_{d-h+1}}= \left( {d\atop d-h+1} \right)=
\left( {d\atop h-1} \right) $
massive degrees of freedom. From the second
expression we can explicitly see  that in this phase it  is the field
$\phi_{h-1}$, rather than $\dphi$, that gets
a mass $m=e{\tilde\Lambda}/2$. Therefore we have {\it three} pairs
of dual actions describing three phases of each of the two
dual theories. We will identify the phase described by
\gleea\ --\daphi\ as `the Coulomb phase', the one described by
\eafda\ --\dafd\ as `the confinement phase' and that described by
\deafda\ as `the Higgs phase'.

For the particular case,
$h=(d+1)/2$ the Higgs and
confinement phases are described by the same type of tensors, since
$\omega_h$ and $\rho_{d-h+1}$ have the same rank.
In this case also the two dual types of topological defects have the
same
dimensionality, since $d-h-1=h-2$. Although we cannot prove it in our
formalism, one would therefore expect that the condensation dynamics,
embodied by the functions $\Lambda ^2 (e)$ and $\tilde \Lambda ^2
(e)$,
respects the duality $e^2 \leftrightarrow 4/e^2$ of the Coulomb
phase, which
means
\eqn\ducoph{\tilde \Lambda ^2 (e) = \Lambda ^2 \left( 2\over e
\right) \ .}
If this is the case, we can immediately conclude that the whole phase
diagram must be symmetric around the self-dual point $e^2=2$.

There are two ways of studying the condensation dynamics: either to
take
into account also the higher modes, neglected in the low-energy
effective
theory, or else to rely on lattice calculations. The structure of the
phase
diagram obtained in these lattice analyses for $d=1,3$ is as follows
\refs{\eli,\cardy}: there
is a confinement phase at strong coupling ($e^2 \gg 2$), dual to a
Higgs phase at weak coupling ($e^2 \ll 2$); in between there is a
self-dual, massless Coulomb phase. Either the Higgs phase or the
confinement
phase disappear if  only one type of topological
defect is taken  into account.

Note also that in the case $h=(d+1)/2$ one can add to the original
low-energy theory a topological term as in \tote . This would highly
increase the complexity of the phase diagram, leading also to
{\it oblique confinement} phases and $SL(2,Z)$ duality \cardy.

Our results can be easily generalized to field theories of several,
coupled
antisymmetric tensors. Besides the fact that these couplings
appear  naturally in string theories,  this is
also  of interest in view of the fact that in several
cases low-energy fermions  can
be represented in terms of antisymmetric tensors, a procedure that
goes
by the name of higher-dimensional bosonization \lut\ and which is of
relevance
both to particle and condensed matter physics \fro.

For example, we could consider a theory containing an $(h-1)$-form
$\phi _{h-1}$ and a $(d-h+1)$-form $\psi _{d-h+1}$ coupled by a
generalized BF-term $\phi_{h-1} \wedge d\psi _{d-h+1}$. In this case
$\psi _{d-h+1}$ plays the same role as the field $V$ in the compact
QED
example of section 4, with the only difference that its dynamics is
specified directly in terms of the kinetic term $d\psi _{d-h+1}
\wedge
\left( d\psi _{d-h+1} \right) ^*$. In the confinement phase for
$\phi _{h-1}$, driven by the condensation of $(d-h-1)$-branes, the
low-energy
induced action for $\psi _{d-h+1}$ describes a Higgs phase and
viceversa.

As a concrete example we would like to mention the (2+1)-dimensional
theory with the Lagrangian
\eqn\mix{{\cal L} = -{1\over 4e^2} F_{\mu \nu }F^{\mu \nu } + {\kappa
\over 2\pi } A_{\mu }\epsilon ^{\mu \alpha \nu } \partial _{\alpha }
B_{\nu } - {1\over 4g^2} f_{\mu \nu }f^{\mu \nu } \ ,}
where $F_{\mu \nu } = \partial _{\mu }A_{\nu }-\partial_{\nu }A_{\mu
}$ and
$f_{\mu \nu }=\partial_{\mu }B_{\nu }-\partial _{\nu }B_{\mu }$.
This theory has  recently  been the focus of many investigations in
connection
with planar strongly correlated electron systems   and Josephson
junction
arrays \mavro. Our results for this theory are in accordance with the
lattice
analysis of ref. \dst.

Let us finish this section with the following remark:
we know that in the limit $\Lambda\rightarrow 0$, the pair of
equations
\eafda\ --\dafd\ reduces to the dual pair \gleea\ --\daphi. The same
can
be said for the pair \deafda\ in the limit
${\tilde\Lambda}\rightarrow 0$.
In both limits the corresponding masses vanish and we recover the
Coulomb phase. It is curious to note that in the confinement phase,
the mass also vanishes if we take the
(strong coupling) limit $e\rightarrow\infty$; however in this case
we {\it do not} recover the original Coulomb phase for the
field $\phi_{h-1}$. We instead obtain a Coulomb phase for the higher
rank
field $\omega_h$, with the coupling constant determined
by  $\Lambda$, as can easily be seen from equations \gffd\ and\eafda.
On the other hand, the dual version  \dafd, will describe in this
limit
a massless field of rank $d-h-1$ dual to the massless field
$\omega_h$.
Similarly, in the Higgs phase, the mass vanishes  in the
limit $e\rightarrow 0$, which gives the Coulomb phase for an
antisymmetric tensor of rank $h-2$ and its dual, of rank $d-h+1$
with coupling given by $\tilde \Lambda$. Therefore,
for a given space-time dimension $D=d+1$,  we can see
that the different phases of antisymmetric tensor theories of any
rank
$r=0,1,\cdots , d$
\foot{The Coulomb phase for a rank $d$ antisymmetric tensor
is only given by a  cosmological constant term, since such a
massless tensor does not have propagating degrees of freedom.} may
{\it all} be connected by changing the different parameters
of each theory, as long as there are topological defects that can
condense.

\newsec{Examples}

We will discuss now some particular examples that illustrate our
generic
results.
 %\item{(i)}
We have already seen in section 4 that the case $d=3,\, h=2$
corresponds to  compact QED. Since this is a case for which
$h=(d+1)/2$, the Higgs and confinement phases can both be described
in terms of a massive vector, dual to
a massive two-index tensor. The massive vector can be either a
`magnetic' or `electric' massive photon. We will now describe some
other examples, probably less familiar.

\subsec{The puzzle of the axion mass}

Let us consider the case $d=h=3$. In this case $\phi _{h-1}$ is the
standard
two-index tensor of string theory $B_{\mu \nu}$
\kalb\  and $\tilde \phi _{d-h}$
is the pseudoscalar axion field $a$ \peccei. The two dual
formulations of the
Coulomb phase are well understood in terms of a single {\it massless}
degree of freedom, implied by the existence of a Peccei--Quinn
symmetry
\peccei:
$a\to a + {\rm constant}$ in the dual theory:
\eqn\dufoax{\eqalign{S &= \int d^4x \ {3\over f^2} \left( \partial
_{[\mu }
B_{\nu \alpha ]} - K_{\mu \nu \alpha } \right) \left( \partial ^{[
\mu }
B^{\nu \alpha ]} - K^{\mu \nu \alpha } \right) \ ,\cr
\tilde S &= \int d^4x \ {1\over 2} \partial _{\mu }a \partial ^{\mu
}a
-a {1\over 16 \pi ^2 f} {\rm Tr} F_{\mu \nu } {F^{\mu \nu }}^* \ ,\cr
}}
where we have included the coupling to the QCD sector, $f$ is a
mass parameter \foot{Note that here we chose $B_{\mu \nu }$ to have
canonical dimension $({\rm mass})^2$.} and $K_{\mu \nu \alpha }$ is
the
dual of the Chern--Simons current, defined by
\eqn\defk{\epsilon ^{\mu \nu \alpha \beta} \partial _{\mu } K_{\nu
\alpha
\beta} = -{1\over 16\pi ^2} {\rm Tr} F_{\mu \nu } {F^{\mu \nu }}^* \
.}
The dual formulation $\tilde S$ was originally introduced to solve
the
strong CP problem; it was immediately recognized that the QCD
instantons
generate a potential $V(a)$, thereby giving the axion a mass. In the
dilute
instanton gas approximation, in which one considers only a dilute gas
of
pointlike instantons, this potential is easily computed to be
$V(a)=\Lambda ^4 \left( 1-{\rm cos}(a/f) \right)$, where $\Lambda ^4$
is the average density of instantons in 4D Euclidean space. The axion
mass in this
formulation is thus $m_a = \Lambda ^2/f$.

Up until recently the origin and description of the axion mass in the
formulation $S$ were a puzzle. This was disturbing since it is
exactly this
formulation of the axion which is obtained in 4D string models.
Last year two independent investigations solved this problem. In
ref.\klls\
it was pointed out that the QCD instantons do indeed generate a mass
for the
$B_{\mu \nu }$ field via the Polyakov mechanism,
but not making it massive itself (a massive two-index
tensor in 4D has three degrees of freedom and cannot be equivalent to
a
massive (pseudo)scalar). The authors concluded that also in this
formulation
there is a physical massive particle without spin but they could not
write an
effective action describing this degree of freedom.
Nevertheless they were still able to find the short-range correlation
function for the field $\epsilon ^{\mu \nu \alpha \beta} \partial
_{[\nu }
B_{\alpha \beta ]}$.
In ref.\bdqq\ the same question
was approached by investigating gaugino condensation in a
supersymmetric version of the string model $S$. It was found that in
such a model the massive
axion must be described by a massive {\it three-index antisymmetric
tensor}
$H_{\mu \nu \alpha }$ which in 4D has one degree of freedom.
The correlation function found in ref.\klls\ corresponds to the
propagator for the dual field
$\epsilon ^{\mu \nu \alpha \beta }H_{\nu \alpha \beta}$.

Let us now explain how these two aspects are unified  by the results
obtained
in the present paper. In the dilute instanton gas approximation the
actions
in \dufoax \ take exactly the form  of the dual actions \extode \ and
\duextode , with $\tilde j_2 = \tilde V_1=0$ and with the instanton
density
$-(1/16 \pi ^2){\rm Tr} F_{\mu \nu }{F^{\mu \nu}}^*$ playing the role
of
$j_0$ and the Chern--Simons term $K_{\mu \nu \alpha }$ playing the
role of
$V_3$. Note that, in this approximation, the QCD instantons get
identified
with the topological defects of a 4D two-index antisymmetric tensor,
the
so-called {\it axionic instantons} \sjt. Using our results we can
immediately conclude that a condensation of these instantons drives a
transition to
a confinement phase, in which  the low-energy action is written in
terms
of a three-index antisymmetric tensor:
\eqn\tiat{S_{-1}= \int d^4x\ \left\{-{3\over 4\Lambda ^4} \partial
_{[\mu }
H_{\nu \alpha \beta ]} \partial ^{[\mu } H^{\nu \alpha \beta ]} +
{3\over f^2} H_{\mu \alpha \beta}H^{\mu \alpha \beta}\right\} \ .}
This action describes one massive degree of freedom of mass $m_H=
\Lambda ^2/f$
dual to the massive axion:
\eqn\masax{\tilde S_{-1}=\int d^4x \ \left\{{1\over 2}\partial _{\mu
}a
\partial ^{\mu }a - {\Lambda ^4\over 2f^2}a^2 \right\}\ .}

Thus it is indeed the condensation of instantons
that generates a mass in the string formulation, as pointed out in
ref.\klls.
And the massive phase must indeed be formulated in terms
of a three-index antisymmetric tensor, as pointed out in ref.\bdqq.
The circle is
closed when one realizes that gaugino condensation is also expected
to be
driven by a condensation of instantons. These results seem  to
indicate
also the possibility to formulate a supersymmetric version of the
Julia--Toulouse mechanism.

Notice that this example goes beyond our prescription in two ways.
First,  the two dual effective actions \tiat\ and \masax, can be
explicitly {\it derived}
in the condensing phase (see the appendix for a sketch of the
derivation in the gaugino condensation case).
Secondly, they do agree with our prescription but only in the
approximation of small scalar field,  for which the $\cos a/f$
potential
reduces to the quadratic term. From the combination of Julia-Toulouse
mechanism and duality, we could only arrive at the mass term for the
axion missing
the fact that since $a$ is a periodic variable, the potential should
also be periodic. We may trace this deficiency in providing the full
scalar potential either to the fact that the Julia--Toulouse
prescription gives only the Lagrangian
up to two derivatives in the fields or to the corresponding
duality between a massive antisymmetric tensor and a massive scalar
which can be performed when the path integrals are Gaussian
(see however the appendix).

Let us finally mention the  Higgs phase ($d=3, h=1$). This would
correspond to the condensation of one-dimensional
objects, strings or vortices. In this case the $B_{\mu\nu}$ field
itself acquires a mass,
dual to a massive vector, each carrying three degrees of freedom,
just as in the compact QED case.
\foot{See the comments at the end of the previous section for a
relation among this phase and the Higgs-confinement phase of compact
QED.} This phase was
also explored in the past by studying in detail the condensation of
strings
in 4D string theory \sjo. The end result was identical to ours.

\subsec{Compact QED and $O(2)$ models in 3D}

%\item{(iii)}
Three-dimensional  QED corresponds to $d=h=2$, dual to
a massless scalar (compact $O(2)$ model).
  The relevant topological defects are instantons
($d-h-1=-1$). Their condensation would generate the confinement
phase described by a massive two-index tensor
carrying one degree of freedom
in 3D. The Higgs phase is obtained by the condensation of
monopoles ($h-2=0$, vortices from the 4D point of view)
and it is described by a massive vector,
the massive photon, dual to a massive scalar,
 coinciding with the results of Polyakov in 3D \poly.
It is interesting to mention that in this case, it has been
shown \poly, that the system is never in a Coulomb
phase. Furthermore, this is also the case for  the $B_{\mu\nu}$
field in 4D \poly\ and for any case in which $h=d$ \orland. This
suggests that
 the 4D axion (and then the string dilaton too) might always be in a
massive phase, although additional couplings might play
an important role.

As we mentioned at the end of the previous section, a massless
limit in the confinement phase would give rise to a massless
two-index tensor which has no dynamics. This completes all
the possibilities for 3D.

\subsec{Strings on a Circle}

This is the $d=h=1$ case. The Coulomb phase is a free 2D scalar,
therefore we can see that this
is relevant for the worldsheet action of string theory:
\eqn\wsh{S={1\over 4\pi \alpha'}\int d^2z\left\{\left(
G_{MN}+B_{MN}\right)\partial_\mu X^M
{\partial^\mu} X^N+\cdots \right\}.}
Where $X^M$, $M=1,\cdots ,D$ are the coordinates of the
$D$ dimensional target space, $G_{MN}$ is the metric in target space,
$B_{MN}$ the antisymmetric tensor and $\alpha'$ is
the inverse string tension. If both are constant we
have the standard $T$ duality $(G_{MN}+B_{MN})\rightarrow (
G_{MN}+B_{MN})^{-1} $\gpr. The case of our interest is the
compactification on a circle of radius $R$.
For which the action is:
\eqn\wsheet{S={1\over 4\pi\alpha'}\int d^2z\,\left\{  \partial_\mu X
{\partial^\mu} X +\cdots\right\}}
Where $X$ is the coordinate of the circle which is
identified with $X + 2\pi R$\foot{
Notice that we are absorbing the coupling constant $1/R$ into the
definition of the
field, unlike the previous sections; this is why $R$ now appears in
the
periodicity conditions. We change our conventions here because
the variable $X$, rather than $\Theta\equiv X/R$, is the standard in
string theory.}  and $R$ is the radius of the circle,
the ellipsis refer to the extra coordinates of the string, as in
\wsh,
which play no role in our discussion.
Duality in the
`Coulomb' phase amounts to the famous  $R\leftrightarrow \alpha'/R$
duality
of the 2D action. We will write the dual action as:
\eqn\wsheetd{\tilde S={1\over 4\pi\alpha'}\int d^2z\,\left\{
\partial_\mu\tilde X
{\partial^\mu}\tilde X +\cdots\right\}}
Where now $\tilde X$ is identified with $\tilde X+2\pi p\alpha'/R,\,
p\in\IZ$. The integer parameter $p$ arises from the Dirac
quantization condition \quant\ between the charge unit of the
topological defects and their duals, as can be easily seen by
comparing with
\dirac\ and with lattice formulations of this model \cardy.
If only one type of topological defect is taken into account,  $p$ is
irrelevant and can just be absorbed into $R$. Instead, it plays a
crucial role if
both types of instantons are taken into account \us.
Notice that $p$ appears in the periodicity condition for the
dual variable and so it corresponds to only  allowing quantized
momenta that are multples of $p/R$.
One-loop modular invariance in string theory would
then require that also winding states should be restricted
indicating that the radius $R$ should be redefined by $R\leftarrow
R/p$
and the effective result would reduce to the $p=1$ case.
We will present the general case for arbitrary $p$ below, keeping
in mind that the only modular invariant string case would be $p=1$.

This is also a case
in which $h=(d+1)/2$. The confinement and Higgs phases are both
driven
by the condensation of instantons and both can be described either in
terms of a massive scalar or  a massive vector. For simplicity we
shall adopt the former description.

Following the terminology introduced in section 5, we call the phase
driven by the condensation of $X$-instantons the confinement phase.
In this phase, the low-energy action in the scalar formulation must
be written
in terms of the dual coordinate $\tilde X$:
\eqn\wshfdd{\tilde{S}_{-1}={1\over 4\pi\alpha'}\int d^2z
\left\{\partial_\mu\tilde X
{\partial^\mu}\tilde X - {\Lambda^2 R^2\over\alpha'^2} \tilde{X}^2
+\cdots\right\}\ ,}
although we expect that, similar to the axion case in 4D, the
quadratic potential in
\wshfdd\ will complete to the periodic form:
\eqn\sgor{
V(\tilde X)={\Lambda^2}\left(1-\cos\left({R\over\alpha'}\tilde
X\right)\right).}
This is the well known sine-Gordon model.
The mass term moves this phase away from a conformal field theory:
the new
phase is confining in the sense that the map from the worldsheet to
target space collapses to a singular map in which antipodal target
space
points are  mapped into a single worldsheet point (in the compact
dimension),
although this is best seen in the massive vector formulation with a
computation similar to the one for compact QED in section 4.
Thus we could say that it is the target space coordinates themselves
that get confined: any space-time interpretation is no longer
possible since the compact coordinate simply disappears.

Contrary to previous examples, in this case
the phase transition point is known.
Indeed the condensation of instantons corresponds to the famous
Berezinsky--Kosterlitz--Thouless (BKT) phase transition
for which the renormalization group flow has been computed
analytically
\refs{\kt,\itz}.
Adapting the BKT results to our notation we find that the transition
occurs at $R=R_c\equiv 2\sqrt{\alpha'}$.
For $R>R_c$ we have the Coulomb (conformal) phase;
for $R<R_c$ we have the confinement phase.
This result, and the corresponding interpretation of $R_c$ as a
minimum radius for compactified strings were already obtained in ref.
\sath\
and confirmed by a matrix model computation in ref. \gk. In addition
to their results, we can
provide an action for the confining phase: note that the
corresponding interpretation of the confining mechanism as the
disappearance of a dimension
is in agreement with ref. \gk, who interpreted it as a reduction
in one unit of the central charge. Notice also that, if $X$ is
originally
a time coordinate, then $R$ would be an inverse temperature: the
phase transition would be a high temperature transition with
critical temperature coinciding with the Hagedorn temperature \sath,
\fintemp.

The existence of a minimum radius was interpreted as a breakdown of
$T$-duality.
As possible ways out, it was suggested a mechanism to discard the
vortices
(instantons for us) \gk, or that another conformal phase might exist
for
$R<\sqrt{\alpha'}/2$ \sath. The solution is actually another.
Indeed, up to now, we have neglected the dual $\tilde X$-instantons.
A condensation of these topological defects drives in fact a
transition to a Higgs phase with action:
\eqn\wshfdd{\tilde{S}_{-1}={1\over 4\pi\alpha'}\int d^2z
\left\{\partial_\mu X
{\partial^\mu} X - {\tilde{\Lambda}^2 \alpha'\over R^2} {X}^2
+\cdots\right\}\ ,}
formulated in terms of the original coordinate $X$.
Following Polyakov \polyt, we interpret the short-range correlations
$\langle\partial X\partial X\rangle$ in this phase as an indication
that the
compact dimension is crumpled. Therefore, also in the Higgs phase we
lose
the space-time interpretation of the compact coordinate, but for a
different
reason.

In addition, the presence of the dual instantons also changes
dramatically
the phase diagram. The new renormalization group flow has been
computed in \refs{\eli, \cardy}. Adapting their results to our
notation we find the following phase structure:
\eqn\nephst{\eqalign{p < 4 &\to \cases{{R^2\over p\alpha '}<1\ ,
& confinement phase\ ,\cr
{R^2\over p\alpha '} >1\ , & Higgs phase\ ,\cr } \cr
p > 4 &\to  \cases{{R^2\over p\alpha '} < {4\over p} \ , &
confinement phase
\ ,\cr
{4\over p}<{R^2\over p\alpha '}<{p\over 4} \ ,
& Coulomb (conformal) phase\ ,\cr
{R^2\over p\alpha '} > {p\over 4}\ , & Higgs phase\ .\cr } \cr }}
Note that the familiar $R\to \alpha '/R$ duality
in absence of topological defects is changed to a $R\to p\alpha '/R$
duality
which is realized as follows. Only the Coulomb (conformal) phase is
self-dual;
the Higgs and confinement phases are instead interchanged under the
duality
transformation. The Coulomb (conformal) phase exists only for $p>4$.
In this case we have both a minimal and a maximal radius
\eqn\minmaxr{\eqalign{R_{\rm min} &= 2\sqrt{\alpha '} \ ,\cr
R_{\rm max} &= {p\over 2} \sqrt{\alpha '} \ .\cr }}
A space-time interpretation is possible only for $R_{\rm
min}<R<R_{\rm max}$.
As we mentioned before, only the $p=1$ case is modular invariant
therefore
modular invariant bosonic strings live only in the Higgs and
confinement
phases which are dual to each other.
Note that these results can be generalized
to the case   of several compact dimensions. For a torus of dimension
$n$, the modular group is $O(n,n;\IZ)$ \gpr\  and the phase diagram
becomes much more complex with the possibility of several conformal
windows
and oblique confinement, as found by Cardy
 for the $SL(2,\IZ)$ case \cardy.
In case the compact coordinate is the time coordinate the maximal
radius
would correspond to a minimal temperature dual to the Hagedorn
temperature.
An extension of these results to the heterotic string is under
current
investigation \us.

\subsec{Higher-dimensional generalizations}

For 10D string theory, an interesting case would be $d=9, h=3$.
Topological defects have dimension $d-h-1=5$ which are usually called
five-branes \dkl. We claim that their condensation will give rise to
a
new phase, described by a  massive
three-index tensor dual to a massive six-index tensor in 10D.
The Higgs phase would  again  be generated by the condensation of
strings
which give a mass to the two-index field dual to a massive
seven-index field in 10D. These are the extensions to 10D
of the axionic instanton results in 4D. The Higgs and
confinement phases have not been previously studied.

 The existence of five-brane solitons in string theory
has been explicitly shown by using the low-energy effective action.
This can easily be generalized to any dimension $D$ starting with
an  antisymmetric tensor $A_{M_1M_2\cdots M_{h-1}}$,
the metric $G_{MN}$ and the dilaton field $\Phi$, with the effective
action
\dkl:
\eqn\effac{S=\int d^Dx\sqrt{-G}\left(R-{1\over 2}\left(\partial
\Phi\right)^2
-{1\over 2h!}e^{-a(h)\Phi} F_h^2\right),}
where $R$ is the curvature scalar, $a(h)$ a constant and
$F_h=dA_{h-1}$.
Notice that the $A_{h-1}$ dependence of this action is only through
$F_h$
and therefore it can be dualized. This action
has regular solitonic solutions of dimension $d-h-1$ and its
dual has regular solitonic solutions of dimension $h-2$. Therefore
it provides us with explicit examples in any dimension, where the
topological defects assumed
previously are present and, if they condense,  we have all the
different phases mentioned before.\foot{An interesting case
is  11D supergravity (effective theory of a yet
unknown $M$-theory): the bosonic spectrum consists of the
metric and a three-index tensor (h=4). This tensor may lead to the
condensation of five--branes. The confinement phase would correspond
to  massive
four-index tensors.
It is believed that due to the existence of a Chern--Simons
like coupling, there is no dual version of this theory in terms of
massless six-index tensors, so the Higgs phase may not be well
defined.} The Higgs and confinement
phases of these theories have not been  considered previously.

Actually a word of care is due at this point. In our previous
examples
all the massless degrees of freedom of the Coulomb phase are
described by
just one single antisymmetric tensor. The situation is different in
\effac , which contains in addition also a massless dilaton and a
massless graviton. Extending our results to this case we can predict
the
antisymmetric tensor content of the Higgs and confinement phases.
However
it is also crucial to know the fate of the dilaton and the graviton
in
these phases.  Supersymmetry will probably help answering these
questions,
which are under current investigation. Various possibilities are
discussed
in the next section.

\newsec{Final Remarks}

 We would like to mention the possible relevance of the present
discussion
to string theory.
We know that string theory has two main problems, namely
how to break supersymmetry and how to lift the
large vacuum degeneracy, especially due to the existence of fields,
such as the
dilaton, that have flat potentials to all orders in perturbation
theory.
The increasing evidence for the existence of a strong--weak coupling
duality
in string theory   has raised the hope that the answer to these
questions may well
be within reach \refs{\sdual,\polch}. There is mounting evidence that
all string theories are related by such a strong--weak coupling
duality  so, strong coupling
effects can in principle be understood by knowing
weak coupling string theory. Even though this is a great step
forward,
it cannot be the full story. The question is that if the
strong coupling domain of  a string theory is determined by the weak
coupling of a different string, and this is given by string
perturbation theory, the problems mentioned above will not
be solved because they were unsolved in perturbation theory.

We hope that the `new' phases we are describing here could
provide a new insight into these questions.
The reason being that if in the confinement and Higgs phases,
the antisymmetric tensors get a mass, then by supersymmetry
also the dilaton will get a mass, since the dilaton is always
in a supersymmetric multiplet with the antisymmetric tensor
$B_{\mu\nu}$; thus the dilaton vev seems to be fixed in
these phases as we wanted, or supersymmetry is broken,
which is also well taken, or both, which may be even better!
For many cases the graviton is also in a multiplet
with antisymmetric tensors and if supersymmetry
is unbroken then the graviton itself will get  a
mass, breaking invariance under general coordinate transformations!
Nevertheless, in the phenomenologically interesting case
of 4D $N=1$ supersymmetry, the graviton belongs to a different
multiplet and we may have massive antisymmetric tensors
with massless gravitons. %does not have to get  a mass.
Still, it will be crucial to understand under which circumstances
supersymmetry is broken in the condensing phases,
or if it is possible at all.

{}From our results and by analogy with the different examples we
studied
in the last section, we may interpret the recently found strong--weak
coupling dualities
among the different 10D string theories, as relations among
the different Coulomb phases of a fundamental theory,
with many confining and Higgs phases in between,
described by massive antisymmetric tensors.
Furthermore, due to the observation at the end of section 5:
for a given dimension, we can start with any massless antisymmetric
tensor
theory and reproduce all the phases of all the antisymmetric tensors
of arbitrary rank. This may be consistent with the
recent claims about $p$-brane democracy    \paul.

As we  already  mentioned, we cannot be very specific about these
issues yet. We can say nothing about the dynamics that
causes the condensation of the topological defects, nor
we can deduce yet the complete effective theory in the new phases.
Nevertheless we believe that  the unified view we are
providing in our discussion and the qualitative description
of the phases, may be a good starting point to
understand the dynamics of these theories and  the complete
phase structure. This may turn out to be crucial in solving the
outstanding questions of string theory. Probably, the recent
developments in terms of Dirichlet-branes \joe\ may provide a
useful tool towards a more concrete investigation of the
process of condensation of the $(d-h-1)$-branes.
It would also be very interesting to find a supersymmetric
generalization
to the Julia--Toulouse mechanism \foot{The results of refs.
\refs{\bdqq,\bin}\
may  be already  a step forward for the 4D case.} so that the
supersymmetry-breaking
question
could be approached in a more quantitative fashion.

\bigbreak\bigskip\bigskip\centerline{{\bf Acknowledgements}}\nobreak
\bigskip
We acknowledge conversations with E. Abdalla, C.P. Burgess, J.
Edelstein,
L.E. Ib\'a\~nez, R. Khuri, R. Minasian, H.P. Nilles and S.-J. Rey.
We also thank P. Townsend for pointing out reference \orland.
The research of C.A.T. was supported by a Profil 2 fellowship of the
Swiss National Science Foundation.

\appendix{A}{}
\noindent{\it Gaugino condensation and duality}
\bigskip

Since the process  of gaugino condensation in $N=1$ supersymmetric
theories
is relatively simple to describe, it is possible in this case to {\it
derive}
the effective action in the confinement phase.
%\foot{Notice that our prescription provides the most general action
% for the %massive antisymmetric tensors up to two derivatives,
% dualization
%gives only  a mass term for the dual field. More general
% `potentials' for the %dual field, such as the $\cos a$ known to be
% induced
%by instantons in the axion formulation, would appear
%only after higher derivative terms are included for the dual field.
%What is a derivative expansion in one action is a small field
% expansion in the
%dual.}.
 This effect is known to be triggered by the
existence of gauge field instantons breaking the
Peccei-Quinn symmetry and therefore the effective theory below
condensation scale coincides with the `confinement' phase
with finite density of instantons. For the sake of
completeness we will now sketch
the main steps of this  derivation.

In 4D, $N=1$ supersymmetric strings, the antisymmetric tensor
belongs, together with the dilaton
and the dilatino,  to a
linear superfield $L$ defined by the constraint $\overline{\cal
DD}L=0$.
This constraint, when expressed in components, implies
the symmetry $B_{\mu\nu}\rightarrow B_{\mu\nu}+$ a closed two-form.
 The simplest scenario to study gaugino condensation
is to consider the couplings of $L$ to a gauge multiplet
of a non-Abelian gauge group $G$ in global supersymmetry. The most
general action is then
the $D$-term of an arbitrary function $\Phi(L)$:
\eqn\linact{
{\cal L}_L =\int d^4x \Dterm{ \Phi(\hat L)}\,}
 with $\hat L\equiv
L-\Omega$ and $\Omega$ the Chern--Simons superfield,
satisfying $\overline{\cal DD}\Omega=\WW$,  $W_\alpha$ is the
gauge field strength superfield, containing in its components,
the gauginos $\lambda_\alpha$ and the gauge field strength
$F_{\mu\nu}$. Expressing this action in components implies,
for instance, that the gauge coupling is given by
$\partial \Phi/\partial L$ and the field $L$ then provides the
field dependent gauge coupling, usual in string theory.
The action \linact\ can be obtained from a first-order
(master) Lagrangian, which can be seen as the supersymmetric
generalization
of \maaphi:
\eqn\mastrss{
{\cal L}(V,S)=
\Dterm{\Phi(V)}
+\Fterm{S\overline{\cal DD}(V +\Omega)},}
where $V$ is an arbitrary vector superfield and $S$ a Lagrange
multiplier chiral superfield ($\overline{\cal D}S=0$).
The components of $S$ are the dilaton, the axion $a$ and their
fermionic superpartner.
Integrating out  $S$, implies
$\overline{\cal DD}(V+\Omega)=0$ or $V=L-\Omega\equiv\hat L$,
giving back the original theory. On the other hand,
integrating first $V$
gives the dual theory in terms of $S$ and $A$ (the gauge superfield).
This is the
situation above the condensation scale.

If the gauge group is asymptotically free, it is expected that
at lower energies the gauge coupling becomes stronger and at a
given scale (the renormalization group invariant scale of $G$),
the gauginos may condense
$\langle \lambda_\alpha\, \lambda_\alpha\rangle\neq 0$. In the
supersymmetric language this is equivalent to requiring
$\langle$Tr$\WW \rangle\neq 0$.
To investigate if condensation takes place we have to construct the
effective action for $\langle$ Tr$\WW \rangle$. We choose to do it
in the first order Lagrangian \mastrss.
We couple an external current $J$ to the operator we want the
expectation value, namely Tr$\WW$:

\eqn\genfun{\exp \left\{{i {\cal W}(J)}\right\}=\int DA\, DS\, DV\,
\exp\left\{i\,
\int d^4x\; \left(
 {\cal L}(V,S)
 +\Fterm{J\WW}\right)\right\}.}
Following the standard effective action procedure \vy,
 we first define the {\it classical} field
$U\equiv \d{\cal W}/\d J=\langle$ Tr$\WW \rangle$.
Integrating first the gauge field $A$, the effective action is a
function of the other variables ($S,V$) and the classical superfield
$U$, $\Gamma(U,V,S)\equiv {\cal W}-\int d^4x\Fterm{UJ}$.
The important result is that since $\cal W$ depends on
$S$ and $J$ only through the combination $S+J$, we can see that
 $\d \Gamma/\d S=\d {\cal W}/\d S=\d {\cal W}/\d J=U$ so
$\Gamma(U,S,V)=US+\Xi(U,V)$, where $\Xi(U,V)$ can be fixed by
the symmetry $U\rightarrow e^{i\alpha}U, S\rightarrow S+i\alpha$.
Therefore $S$ appears only
linearly in the path integral and its integration gives
again a $\delta$-function,
imposing now $\overline{\cal{DD}}V=-U$ instead of
the constraint $\overline{\cal DD}(V+\Omega)=0$ above the
condensation scale.
We can then see that since the constraint on $V$ is different,
there is no linear
multiplet implied by this new  constraint. This  is an indication
that
the $B_{\mu\nu}$ field is no longer in the spectrum.

The effective action in components can be easily written \bdqq.
Considering two condensing groups, freezing the dilaton
degree of freedom and eliminating the auxiliary fields by their
field equations, we end up with a Lagrangian of the form:
\eqn\comp{{\cal L}=H_{\mu\nu\rho}^2-a\epsilon^{\mu\nu\rho\sigma}
\partial_\mu H_{\nu\rho\sigma}+(a-\theta)^2 +(1-\cos\theta)
+f(\theta)\partial_\mu \theta\partial^\mu\theta\ ,}
where $\theta$ is the difference in phases of the two
classical fields ($U_1$ and $U_2$) representing the condensate.
The function $f(\theta)$ defines the kinetic term for $\theta$.
The important point here is that $\theta$, unlike $H_{\mu\nu\rho}$
and
$a$
is not a variable to be integrated in the path integral, it is only a
field to be eliminated by
its field equations. We can easily see that integrating out $a$
and setting $\theta$ at the minimum of its potential
$\theta=0$ we obtain:
\eqn\ache{
{\cal L}=H_{\mu\nu\rho}^2-{1\over 4}\left(\partial_\mu
H_{\nu\rho\sigma}\right)^2}
whereas integrating out $H_{\mu\nu\rho}$ and setting $\theta$ at the
minimum of its potential $\theta=a$ we arrive at the
Lagrangian for the axion
\eqn\axion{
{\cal L}=-{1\over 4}\left(\partial a\right)^2+\left(1-\cos a\right).}

\listrefs

\end